\documentclass[aps,graphicx,twocolumn,groupedaddress,superscriptaddress]{revtex4-2}%,linenumbers

\usepackage{CJK}
\usepackage{amssymb}
\usepackage{amsmath}
\usepackage{dcolumn}
\usepackage{multirow}
\usepackage{graphicx}
\usepackage{subfigure}
\usepackage{float}
\usepackage{epstopdf}
\usepackage{mathrsfs}
\usepackage{appendix}
\usepackage{booktabs}
\usepackage{color}
\usepackage[table]{xcolor} 
\usepackage{titlesec}
\usepackage{url}
\usepackage[colorlinks]{hyperref}
\hypersetup{%
	plainpages=true,
	breaklinks=true,%not default in dvips mode, so we must specify
	hypertexnames=false,%not ideal, but needed when pagenums duplicate (`i' vs. `1')
	pageanchor=true,
	colorlinks=true,
	linkcolor={blue},
	citecolor={red},
	urlcolor={blue},
	anchorcolor={black}
}

\makeatletter
\renewcommand{\maketag@@@}[1]{\hbox{\m@th\normalsize\normalfont#1}}%
\makeatother
\allowdisplaybreaks[4]

\begin{document}
%\linenumbers\href{mailto:tao.li@njust.edu.cn}{\ast}
%\email{tao.li@njust.edu.cn}$
	\preprint{APS/123-QED}
\title{Heralded  nonlocal quantum gates for distributed quantum computation \\ in a decoherence-free subspace}
\author{Wanhua Su}
\affiliation{MIIT Key Laboratory of Semiconductor Microstructure and Quantum Sensing, School of Physics,  Nanjing University of Science and Technology, Nanjing {\rm 210094}, China}
\author{Wei Qin}
\affiliation{Center for Joint Quantum Studies and Department of Physics,School of Science, Tianjin University, Tianjin 300350, China}
\affiliation{Theoretical Quantum Physics Laboratory, Cluster for Pioneering Research, RIKEN, Wakoshi, Saitama 351-0198, Japan}
\author{Adam Miranowicz}
\affiliation{Theoretical Quantum Physics Laboratory, Cluster for Pioneering Research, RIKEN, Wakoshi, Saitama 351-0198, Japan}
\affiliation{Institute of Spintronics and Quantum Information, Faculty of Physics  and Astronomy, Adam Mickiewicz University, PL-61-614 Pozna\'n, Poland}
\author{Tao Li}
\email[]{tao.li@njust.edu.cn}
\affiliation{MIIT Key Laboratory of Semiconductor Microstructure and Quantum Sensing, School of Physics,  Nanjing University of Science and Technology, Nanjing {\rm 210094}, China}
\affiliation{Engineering Research Center of Semiconductor Device Optoelectronic Hybrid Integration in Jiangsu Province, Nanjing {\rm210094}, China}
\author{Franco Nori}
\affiliation{Theoretical Quantum Physics Laboratory, Cluster for Pioneering Research, RIKEN, Wakoshi, Saitama 351-0198, Japan}
\affiliation{Quantum Information Physics Theory Research Team, Quantum Computing Center, RIKEN, Wako-shi, Saitama 351-0198, Japan}
\affiliation{Physics Department, The University of Michigan, Ann Arbor, Michigan 48109-1040, USA}
\date{\today}% It is always \today, 

\begin{abstract}
We propose a heralded  protocol for implementing nontrivial quantum gates on two stationary qubits coupled to  spatially separated cavities. By dynamically controlling the evolution of the composite system, nonlocal two-qubit quantum (e.g., CPHASE and CNOT) gates  can be achieved without  real excitations of either cavity modes or atoms. The success of our protocol is conditioned on projecting an auxiliary atom  onto a postselected state, which simultaneously removes  various detrimental effects of dissipation on  the gate fidelity. In principle, the success probability of the gate can approach unity as the single-atom cooperativity becomes  sufficiently large.
Furthermore, we show its application for implementing single- and two-qubit gates within a decoherence-free subspace that is immune to  a collective dephasing noise. This faithful, heralded, and nonlocal protocol could, therefore, be useful for distributed quantum computation and scalable quantum networks.
\end{abstract}

\maketitle

	%%%%%%%%%%%%%%%%%%%%%%%%%%%%%%%%%%%%%%
\section{Introduction}
\label{se:introduction}
%\sloppy{}
	
Quantum computation exploiting quantum systems for information processing has attracted a great  {deal of attention}~\cite{ladd2010quantum,slussarenko2019photonic,Kockum2019,buluta2011natural,zhou2020limits,qin2024quantum,devoret2013superconducting, wendin2017quantum,gu2017microwave} due to its promising advantages over classical computation~\cite{Long2006general,Georgescu2014Quantumsimulation,Yuan2020quantum-computing}, and has   {been} experimentally demonstrated   {with} its superiority in handling well-defined
tasks. These include implementing algorithms based on quantum
gates~\cite{Arute2019Quantum, wu2021strong} and quantum
annealing~\cite{King2021Scaling} using superconducting quantum
processors, and performing boson sampling using linear-optical
interferometers~\cite{Zhong2020, Zhong2021Phase-Programmable,
Madsen2022}.
Nontrivial two-qubit quantum gates in combination with general single-qubit rotations in principle enable  {implementing} various quantum algorithms for practical applications. The two-qubit quantum gates always involve direct or indirect interactions between  {the} systems   {which} they are applied on. So far, two-qubit quantum gates have been proposed for different physical systems, such as   {photons} ~\cite{duan2004scalable,koshino2010,reiserer2014quantum,tiecke2014nanophotonic,Nemoto2014Photonic,Liu2020Low-Cost}, trapped ions~\cite{bruzewicz2019trapped,postler2022demonstration}, color centers~\cite{qin2019proposal,Wei2013Compact,Li2016Hybrid,burkard2017designing, chen2019universal,Li2020Enhancing,Zhou2023Parallel}, quantum dots~\cite{Hu2008Giant,Lit2016Error,Piparo2019multiplexing,lodahl2015interfacing}, and  {superconducting circuits}~\cite{you2011atomic,Wei2008Controllable,Guo2023Heralded}. 
However, the scalability of quantum computation is challenging due to the inevitable presence of noise and  decoherence. Fortunately, their influence on the evolution of quantum  systems can be suppressed by   {the use of},  {e.g., dynamical}  decoupling~\cite{viola1999dynamical,pokharel2018demonstration}, holonomic manipulation~\cite{ZANARDI199994,Xu2012Nonadiabatic,Zhao2020General}, and decoherence-free subspaces~(DFSs)~\cite{lidar1998decoherence,kockum2018decoherence,Nathan2018Open,Zhang2019Protection}. Moreover, a certain amount of noise and decoherence can be tolerated by   {harnessing} quantum error-correction  {codes~\cite{terhal2015quantum,Li2024Heralded},} in which the overheads and   {the} complexity considerably increase with the error rate.
%zu2014experimental,PhysRevLett.122.010503,	
	
For	some specific dominant noise or   {decoherence~\cite{lidar1998decoherence}, DFSs} can provide an efficient method for protecting the logical qubits {against noise} by encoding   {quantum information} in  {a} DFS~\cite{Wu2005Holonomic,You2005Correlation,Miccuda2020Decoherence-Resilient,Monz2009Realization,Qiao2022Generation}. A fundamental and dominant noise in stationary systems is dephasing due to   {the} random fluctuations of external fields~\cite{Monz2009Realization}, which   {destroy} the coherence between two computational basis states. A simple DFS for tackling this issue can be constructed by properly encoding a logical qubit with two physical qubits,   {which} simultaneously   {suffers} from the same phase noise (i.e., collective dephasing noise)~\cite{lidar1998decoherence}. Exploiting   {DFS} for quantum computation has been widely   {studied}  {using} {various  platforms}~\cite{xue2006universal,Cen2006Scalable,Brion2007Universal,Deng2007Preparation,
Chen2010Quantum,Song2016Shortcuts,Wu2017Adiabatic,Zwerger_2017,zhang2018holonomic,
Farfurnik2021Single-Shot,Hu2021Optimizing,Sun2022One-step,Chen2020Robus,Chen2022Fault-Tolerant,Du2024Decoherence}. For these protocols,    {a} {DFS} can work in a deterministic way by dynamically controlling the evolution of systems, or in a heralded way with the detection of single photons scattered by  cavity-coupled platforms. Furthermore, some significant experimental efforts have been made for the realization of quantum gates acting on  {decoherence-free systems}~\cite{Feng2013Experimental,	Zhu2019Single-Loop,Xu2020Experimental,Blumoff2022Fast,han2024protecting}.

Recently, a heralded method for achieving effective quantum computation~\cite{borregaard2015heralded,qin2017heralded,kang2020heralded}  {has been} presented by dynamically controlling the evolution rather than by scattering and measuring single photons. Borregaard \textit{et al.}~\cite{borregaard2015heralded} proposed a heralded, near-deterministic protocol for performing quantum gates on natural atoms trapped in a single optical cavity. Qin \textit{et al.}~\cite{qin2017heralded} presented  heralded, controlled-phase (CPHASE) gates on superconducting qubits   {coupled} to the same cavity, and introduced a  {spatially separated} cavity coupled to an auxiliary qubit for {a} heralding operation. These protocols provide a  {\textit{ quadratic  fidelity improvement}} compared to previous deterministic cavity-based gates, and can find their applications in long-distance entanglement distribution and quantum computation~\cite{borregaard2015long,borregaard2017efficient,kang2020heralded,Qin2018Exponentially}. 

However, it is noteworthy that  {nontrivial two-qubit gates applied on spatially separated} stationary qubits coupled to different optical cavities are useful for {connecting several distinct quantum information processors, which constitute the backbone for distributed quantum computation~\cite{Cirac99Distributed,Lim2005Repeat,Jiang2007Distributed,zheng2010arbitrary} and scalable quantum repeater networks~\cite{Briegel98Qrepeater,jiang2009quantum,Wang2012QR,munro2012quantum,sheng2013hybrid,wehner2018quantum,yan2021survey}.} Hence   it is  important  to generalize the heralded schemes of Refs.~\cite{borregaard2015heralded,qin2017heralded} to the {\textit{nonlocal}} case, {where \textit{nontrivial two-qubit quantum gates applied on two spatially separated qubits can be generated in a heralded architecture by dynamically controlling and measuring the auxiliary atom.} For simplicity of notation, we refer to quantum gates applied on spatially separated qubits as \textit{nonlocal gates} when there is no ambiguity.}

In this paper,  {we propose a heralded method for implementing nontrivial quantum gates acting on spatially separated  stationary qubits} coupled to different cavities by dynamically controlling the evolution of cavity-coupled systems. The cavities can be connected by short fibers or superconducting coaxial cables~\cite{reiserer2015cavity}. A four-level auxiliary  {atom}  {is} coupled to an additional cavity as both a virtual-photon source and a detector for heralding the success of the quantum gate~\cite{borregaard2015heralded,qin2017heralded}. According to the results of a proper measurement on the auxiliary  {atom},   {\textit{{the gate} errors introduced by   {atomic spontaneous emission and cavity photon loss} can be inherently removed, leading to faithful two-qubit nonlocal gates.}}  As a result, the detected errors simply lower the success probability of the gate rather than its fidelity, which is extremely important for practical applications~\cite{Cirac99Distributed,Lim2005Repeat,Jiang2007Distributed,zheng2010arbitrary,Briegel98Qrepeater,jiang2009quantum,Wang2012QR,munro2012quantum,sheng2013hybrid,wehner2018quantum,yan2021survey}.

We show that \textit{the fidelity of our nonlocal two-qubit gate can be  further improved  by applying proper single-qubit operations   {to} the qubits before completing  the two-qubit gate.} Furthermore,  we propose an approach for performing a heralded nontrivial two-qubit gate  {in} a DFS immune to collective dephasing noise. Each logical qubit consisting of two physical qubits couples to an individual cavity and suffers from different dephasing noises. Combining the advantages of heralded inherent error detection and error-avoiding DFS, our protocol for implementing nonlocal quantum gates can directly find its applications in distributed quantum computation and quantum networks.

The  {remainder of the} paper is organized as follows: {In Sec.~\ref{se:Hera_CZ},  we describe the physical model and mechanism for implementing a heralded nonlocal two-qubit gate on two spatially separated qubits. In Sec.~\ref{se:master_eq},  we introduce the effective Hamiltonian and Lindblad operators after conditionally 
excluding dissipative quantum jumps. In Sec.~\ref{se:simulation}, we describe an implementation of a heralded nonlocal
CPHASE gate and analyze its performance both analytically, using the effective Hamiltonian and Lindblad operators in Sec.~\ref{se:master_eq}, and numerically through a master equation simulation. In Sec.~\ref{se:DFS}, we present heralded nonlocal two-qubit gates  {operating on} logical qubits  {in} a DFS  immune to collective dephasing noise.} Finally, we conclude with a brief discussion and summary in Sec.~\ref{se:conclusions}.

\begin{figure}[tbp]
	\includegraphics[width=8.8cm]{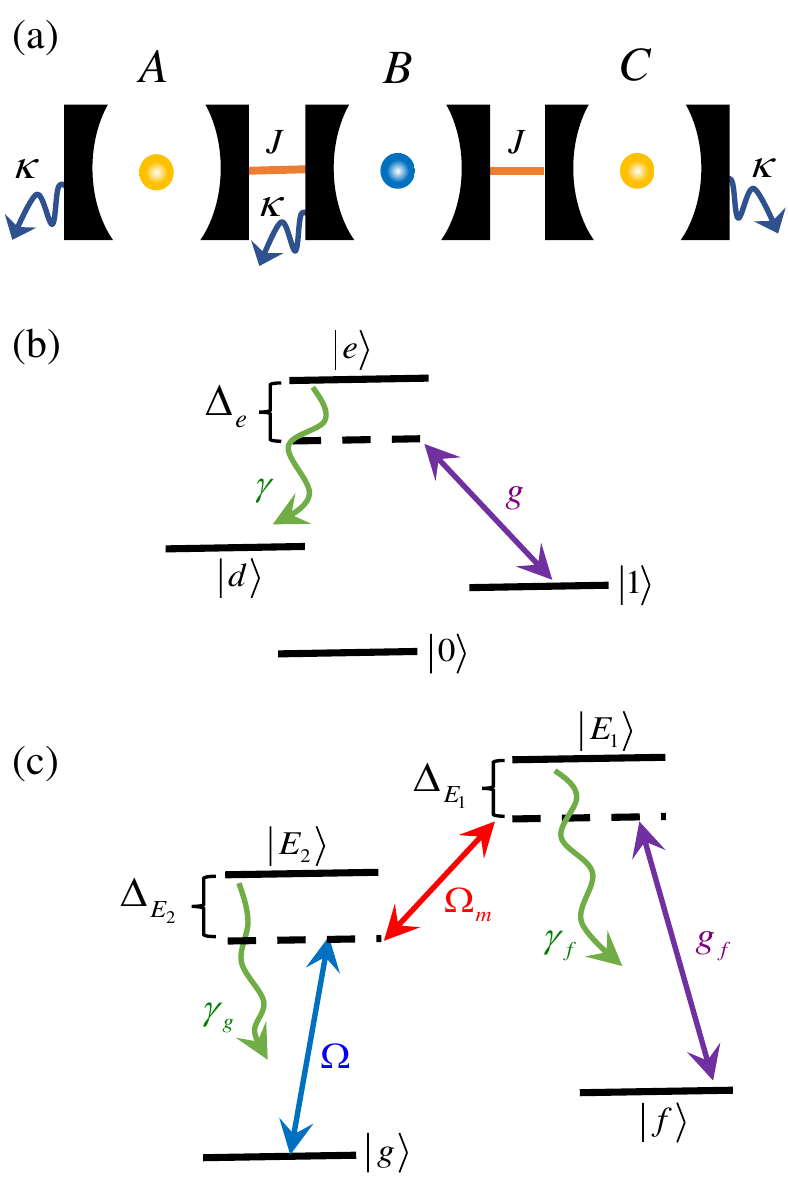}
		\caption{Schematics of a heralded nonlocal two-qubit quantum gate. (a)   {Implementation} of  {the} gate with   {a} cavity-coupled   {system}. Two stationary qubits are distributed in two separated cavities that are connected to an auxiliary cavity via short fibers or superconducting coaxial cables. (b)  Level structure of two {qubit-encoding} atoms coupled to  cavities A and C. (c) Level structure of the auxiliary atom that couples to cavity B and   {works} as a heralding {system.}
		}
		\label{fig1}
	\end{figure}

%\section{Prototype physical model for a heralded entangling gate}
\textcolor{black}{\section{Physical mechanism and configuration for implementing heralded nonlocal two-qubit gates}	\label{se:Hera_CZ}}

An essential building block for implementing heralded nonlocal two-qubit gates is  {the use of} cavity-coupled systems~\cite{reiserer2015cavity}. They can be implemented by various natural or artificial atoms~\cite{buluta2011natural} coupled to optical cavities~(including transmission-line  resonators), which   {can be} connected by short optical fibers~(or superconducting coaxial cables).
	
	{\begin{table}[!ht]
		\caption{ {Basic notations used in this paper.} } \label{tab1} \label{table}
		{\color{black} 
            \begin{tabular}{*{1}{p{2.0cm}<{\centering}}*{1}{p{6.4cm}<{\centering}}}
            			\hline \hline  
			Notation & Meaning\\ [0.5ex] 
			\hline
			$\omega_{x}$ & Frequency of the atomic state $|x\rangle$  \\
			$\omega_{c}$ & Common resonance frequency of the cavities $A$, $B$, and $C$\\
			$\omega_{L}$, $\omega_{m}$ & Frequencies of the classical driving fields \\
			$\Omega_{L}$, $\Omega$ &  Rabi frequencies of the classical driving fields   \\
			$g$ ($g_{f}$) & Coupling strength between the qubit (auxiliary) atom and the cavity  \\
			$J$ & Inter-cavity coupling strength   \\
			$\gamma$,$\gamma_{g}$,$\gamma_{f}$ & Decay rates of atomic excited states  \\
			$\kappa$ & Cavity decay rate    \\
			$C=g^{2}/\left(\kappa\gamma\right)$ & Atom-cavity cooperativity    \\
			$\Delta_{E_1}$, $\Delta_{E_2}$ & Detunings for the one- and two-photon transitions in the auxiliary atom\\
            $\Delta_e$ & Detuning of the qubit-encoding atom from the normal mode $c_1$\\
            $\mathcal{P}_N$ & Operators projecting the qubit-encoding atoms onto a state with $N$ qubits in $\left | 1  \right \rangle$\\
            $\Delta_N$ & $N$-dependent ac Stark shifts\\
            $L_{\rm eff}^\zeta$ & Effective Lindblad operators for $\zeta=$$f$, $g$, $c_l$, and $k$\\
            $r_ {\zeta,{\scriptsize N}}$ & Effective decay rates of $L_{\rm eff}^\zeta$\\[0.5ex] 
			\hline
            \hline
		\end{tabular}
         }
	\end{table}}

The  schematics of our heralded nonlocal protocol
is shown in Fig.~\ref{fig1}.  Two  {qubit-encoding} atoms couple to two   {separated} cavities A and C, respectively, which are connected via short optical fibers, and an auxiliary atom   {couples} to   cavity B in the middle.  The effective coupling   {between} two neighboring cavities can be described by a  coupling rate  $J$ when the fiber length $L$ is small and two cavities are resonant~\cite{cho2008heralded,Serafini2006Distributed}.

{{A \textit{collective} normal mode can be formed as a  linear combination of these cavity modes. It interacts \textit{simultaneously}  with all the atoms,  when all the cavity modes are   resonant and strongly interact  with the neighboring cavity modes through photon exchange. A distributed quantum gate,  operating on spatially separated qubit-encoding atoms, as shown in Fig.~\ref{fig1}(a),  can be simplified to a quantum gate  acting on the atoms coupled to the same cavity mode~\cite{borregaard2015heralded}.}} 

{{Each  {qubit-encoding} atom has two ground levels~($\left | 0  \right \rangle $  and $\left | 1  \right \rangle $), which   {can} encode a qubit, and one excited level $\left | e  \right \rangle $, shown in Fig.~\ref{fig1}(b). We assume that the transition $\left | 1  \right \rangle\leftrightarrow \left | e  \right \rangle$  of both  {qubit-encoding} atoms is coupled to the cavity mode with a coupling rate $g$ and a detuning $\Delta_e$, and that the excited level $\left | e  \right \rangle$ decays to a level $\left | d  \right \rangle$,   {which may or may not be}   {$\left | 0  \right \rangle $ or $\left | 1  \right \rangle $}.}}

{{The auxiliary atom has two ground states ($\left | g  \right \rangle $  and  $\left | f  \right \rangle $) and two excited states ($\left | E_1  \right \rangle $ and $\left | E_2  \right \rangle $), shown in Fig.~\ref{fig1}(c). The excited states  $ \left | E_1  \right \rangle$ and $ \left | E_2  \right \rangle$ spontaneously decay to   {the} ground states $ \left |f  \right \rangle$ and $ \left | g \right \rangle$ with  rates $\gamma_f$ and $\gamma_g$, respectively. In addition, the   $\left | f  \right \rangle\leftrightarrow \left | E_1  \right \rangle$ transition  couples to the cavity mode $a_B$ with a coupling rate $g_f$ and a detuning $\Delta_{E_1}$. The transition between the states $\left | E_2  \right \rangle$ and $ \left | E_1  \right \rangle$ ($\left | g  \right \rangle$ and $ \left | E_2  \right \rangle$) is driven by a classical field with frequency $\omega_{m}$  {($\omega _L$)} and the Rabi frequency $\Omega_m$ ($\Omega$). Therefore, a three-photon resonant transition, resulting in a flip of the two ground states of the auxiliary  {atom}, can be achieved by tuning the driving frequencies $\omega_m$ and $\omega_L$.}} 

{{In general, the auxiliary atom involves three independent transitions from the ground state $|g\rangle$: the single-photon transition  $|g\rangle\leftrightarrow|E_2\rangle$, the two-photon transition $|g\rangle\leftrightarrow|E_1\rangle$, and the three-photon transition $|g\rangle\leftrightarrow|f\rangle$. When all qubit-encoding atoms decouple from the collective mode and the three-photon resonance transition is achieved, the auxiliary atom can  evolve  into a dark zero-energy state after removing the Stark shift of the ground state $|g\rangle$ that is introduced by the nonresonant single-photon transition. Note that the auxiliary atom remains almost unchanged, and the excitation of the collective mode is negligible for weak driving fields.}} 

{{Conversely, when the qubit-encoding atoms couple to the collective mode, the frequency of the collective mode is shifted, and the three-photon resonance condition is no longer satisfied. As a result, the combined state of the system mainly experiences the single-photon and two-photon transitions for weak driving fields with large detunings. The two-photon transition introduces an additional energy shift of the ground state $|g\rangle$, which is nearly independent of the number of the coupled qubit-encoding atoms due to the weak excitation of the collective mode. By appropriately adjusting the driving pulse length, a relative phase shift of $\pi$ can be introduced for the decoupled state of the qubit-encoding atoms compared to the case of all the coupled states.}}

{{The decay of either the atoms or the cavity modes leads to the collapse of the auxiliary atom into the state $|f\rangle$. This collapse can be heralded by measuring the auxiliary atom, other than relying on the null detection of the photons leaving the cavity. By postselecting state $|g\rangle$ of the measurement on the auxiliary atom as a heralding signal, errors introduced by finite decay rates are then converted into a non-unity probability of success. Moreover, the excitations of the cavity modes and the excited states of the atoms are negligible and can be adiabatically eliminated, when the system is initially prepared in the ground-state subspace for weak driving fields and large detunings. Consequently, we can concentrate on the evolution of the ground state and describe the corresponding dynamics using an effective Hamiltonian that excludes the dissipation of atomic and cavity excitations.}}

The {total} Hamiltonian of the composite system, consisting of  {the} three atoms and three cavities, can be   {written} as
 \begin{eqnarray}
H_T=H_0+H_1,
 \end{eqnarray}
where $H_0$ and $H_1$ represent the free and interaction Hamiltonians, respectively. The free  Hamiltonian $H_0$   is
\begin{align}\label{eq:H0}
	{H_0} =&  \sum\limits_{k = 1,2} \big ( {\omega _e{\left| e \right\rangle }_k\left\langle e \right|
		+ \omega _1{\left| 1 \right\rangle }_k\left\langle 1 \right|
		+ \omega _0{\left| 0 \right\rangle }_k\left\langle 0 \right|} \big )\nonumber\\
	&+ \omega _{E_1}\left| E_1 \right\rangle \left\langle E_1 \right|
	+ \omega _{E_2}\left| E_2 \right\rangle \left\langle E_2 \right|
	+ \omega _f\left|f \right\rangle \left\langle f\right|\nonumber\\
	&+ \omega _g\left|g\right\rangle \left\langle g \right|
	+ {\omega _c}( {a_A^\dag {a_A} + a_B^\dag {a_B} + a_C^\dag {a_C}}),
\end{align}
where $\omega_x$ is the frequency of the atomic   {level}  $\left|x\right\rangle$, except  {$\omega_c$, which is} the common resonance frequency of the three cavities.
The interaction Hamiltonian $H_1$  (including the cavity-cavity coupling, the atom-cavity coupling, and the classical driving) {becomes}
 \begin{eqnarray}
	H_{1}&=&\Big[g( {a_A}{{\left| e \right\rangle }_1}\left\langle 1 \right| +{a_C}{{\left| e \right\rangle }_2}\left\langle 1 \right|)
+{g_f}{a_B}\left| {E_1} \right\rangle \left\langle f \right|
	\nonumber\\&&+\frac{1}{2} \left( \Omega{e^{ - i{\omega _L}t}}\left| E_2 \right\rangle \left\langle g \right| +\Omega _{m}{{e^{ - i{\omega _m}t}}\left| {{E_1}} \right\rangle \left\langle {{E_2}} \right| }\right)
		\nonumber\\ &&+	J({a_A}a_B^\dag + {a_C}a_B^\dag)
	\Big]+ \rm{H.c.},
\end{eqnarray}
where \rm{H.c.} represents the  Hermitian conjugate, and we   {have assumed} a symmetric coupling between the two  {qubit-encoding} atoms and their corresponding cavities.

In order to explicitly describe the dynamics of the composite system, we perform a transformation   {for} the three cavity modes and introduce three  delocalized normal modes as:
 \begin{eqnarray}
c_{1} &=&\dfrac{1}{2} ( a_{A}-\sqrt{2}a_{B}+a_{C}  ),
	\nonumber\\
c_{2} &=&\dfrac{1}{2} ( a_{A}+\sqrt{2}a_{B}+a_{C} ),
\nonumber\\
c_3&=&\dfrac{1}{\sqrt{2}} ( a_{A}-a_{C}).
\end{eqnarray}
% $c_{1} = ( a_{A}-\sqrt{2}a_{B}+a_{C}  ) /2$, $c_{2} = ( a_{A}+\sqrt{2}a_{B}+a_{C} ) /2$, and $c_{3} = ( a_{A}-a_{C}  ) /{\sqrt{2}}$.
The   {total} Hamiltonian in the new basis can be described, in a proper rotating frame, as
 \begin{eqnarray}
H_T=H_e+V+V^\dag, \label{HamiltonianT}
\end{eqnarray}
where $H_e$ and $V$ describe the evolution of the single-excitation   {subspace} and its coupling to the ground space, respectively.   {Specifically, they}  {can be   expressed as}
\begin{align}\label{eq:He}
	{H_e} = &{\Delta _{E_1}}\left|{E_1}\right\rangle \left\langle {E_1} \right|
	+{\Delta _{E_2}}\left|{E_2}\right\rangle \left\langle{E_2} \right|\nonumber\\
	&+{\left[ \frac{\Omega _m}{2} \left|{E_1}\right\rangle \left\langle{E_2}\right|+\rm{H.c.} \right]}+H_{e1},
\end{align}
	where
\begin{align}\label{eqHe1}
	H_{e1}=
	& \sum_{k=1,2}
	 \Big\{\Big[ \frac{g}{2}( {{c_1}+{c_2}+\sqrt{2}S_k {c_3}} ){{\left|{e}\right\rangle }_k}\left\langle{1}\right| + \rm{H.c.}  \Big]\nonumber
	 	 \\& +\Delta _e{\left|{e}\right\rangle _k}\left\langle{e}\right|\Big\}
	+{\Big[ \frac{g_f}{\sqrt 2 }\left( {{c_2} - {c_1}}\right)\left| {{E_1}} \right\rangle \left\langle f \right|
		+ \rm{H.c.}  \Big]}\nonumber
		 \\ & {+\sum_{i}^{3}\Lambda_i c_i^\dagger {c_i}},
\end{align}
{with $S_k=(-1)^{k+1}$} ,  {${\Lambda_1 } = {\omega_c} - \sqrt 2 J$,  ${\Lambda_2 }= {\omega_c}+\sqrt{2} J$, ${\Lambda_3 }= {\omega_c}$,}  and $V = \frac{\Omega }{2}\left|{E_2}\right\rangle \left\langle{g}\right|$.
Here, for simplicity, we {have defined} some detunings as follows:
	\begin{eqnarray}
		{\Delta _{E_1}} &=& {\omega _{E_1}} - {\omega _L} - {\omega _m} - {\omega _g}, \nonumber\\
		{\Delta _{E_2}} &=& {\omega _{E_2}} - {\omega _L} - {\omega _g}, \nonumber\\
		{\Delta _e} &=& {\omega _e} - {\omega _L} - {\omega _m} + {\omega _f} - {\omega _g} - {\omega _1}.
	\end{eqnarray}
For large   {detunings ({i.e., }${\Delta _{E_1}\gg\Omega }$ and ${\Delta _{E_2}\gg\Omega _m}$)} and a large coupling  {strength ({i.e.,} $J\gg g_f$) between} two neighboring cavities, we can effectively eliminate the excited states $|E_1\rangle$ and $|E_2\rangle$ and {then} obtain a three-photon resonant Raman transition $\left | {g}  \right \rangle \to\left | {f}  \right \rangle$, which is mediated by mode $c_1$ rather than   modes $c_{2,3}$   {if} the driving field frequency is tuned to
	\begin{align}\label{eq:omegac}
{\omega_{L}} =  	{\omega _c}-{\omega _m}+{\omega _f} - {\omega _g}-\sqrt{2} J,
	\end{align}
i.e. $\Lambda_1 = 0$.
The evolution of the composite system consisting of two  {qubit-encoding} atoms,  {a single} auxiliary  {atom}, and three cavities connected by optical fibers can in principle be identical to that of two  {qubit-encoding} atoms and one auxiliary  {atom,} all directly coupled to the same cavity mode~\cite{borregaard2015heralded}.

By adiabatically eliminating  state $|E_2\rangle$ of the auxiliary  {atom} and moving   {into} a   {proper} rotating frame, the effective Hamiltonian of the composite system can be  {described by} $H'_T={H'_e}+V'+V'^{\dag}$, with an effective three-level auxiliary  atom, 
\begin{align}\label{eq:He2}
	{H'_e} ={\Big(\Delta _{E_1}-\frac{\Omega_m^2}{4\Delta_{E_2}}\Big)}\left|{E_1}\right\rangle \left\langle {E_1} \right|+H_{e1},
\end{align}		
and
\begin{eqnarray}
V'=-\tilde{\Omega}\left|{E_1}\right\rangle\left\langle{g}\right|, \;\;\; \tilde{\Omega}= \frac{\Omega _m\Omega}{2\Delta_{E_2}},
\end{eqnarray}
{where the energy of the ground state $|g\rangle$ has been shifted by $\Omega^2/\left(4 \Delta_{E_2} \right)$, which can be achieved by using a laser that couples to $|g\rangle$ nonresonantly with an additional level.}

When all  {qubit-encoding} atoms are in  state $|0\rangle$ that   {is} decoupled from mode $c_1$, an adiabatic excitation of the auxiliary  {atom results in the dark zero-energy state:}
 \begin{eqnarray}
|\psi\rangle_d=\frac{1}{\sqrt{g_f^2+2\tilde{\Omega}^2}} \left(g_f|0,0,0,g\rangle-\sqrt{2}\tilde{\Omega}|1,0,0,f\rangle\right),\;\;\;\;\;\;\;\;
\end{eqnarray}
where $|0,0,0,g\rangle$ represents {the} three normal modes in the vacuum state and the auxiliary state is $|g\rangle$, while $|1,0,0,f\rangle$ represents that mode $c_1$  {has a single photon,}  modes $c_2$   {and} $c_3$ are in the vacuum state, and the auxiliary  {atom}  is in  state $|f\rangle$. {For weak driving fields with large detunings, the dark state $|\psi\rangle_d$ approaches $|0,0,0,g\rangle$,
and the excitation of the normal modes can be considered negligible with a probability approximately zero, $\left[\Omega_m\Omega/\left(\Delta_{E_2}g \right)\right]^2\sim0$.}

{{In contrast, when either or both  qubit atoms are in state   $|1\rangle$, they couple to mode $c_1$, 
thereby distorting the three-photon resonant condition. This introduces the ac Stark shifts arising from the nonresonant one- and two-photon transitions and leads to dynamical phases upon applying the driving fields.}} Therefore, all  {the} qubit states, except the uncoupled one, acquire a phase that is determined by the duration of the driving field, which is essential for constructing various heralded nonlocal quantum  {gates (as shown below).}

	\textcolor{black}{\section{Effective Hamiltonian and Lindblad operators following the conditional exclusion of dissipative quantum jumps}\label{se:master_eq}}

{So far, we have provided a qualitative description of the physical model and mechanism for the implementation of the heralded nonlocal two-qubit gates; focusing particularly on the ideal scenario, where the composite system remains decoupled from its environment. In this section, we proceed to a quantitative analysis of the physical mechanism, where we derive an effective Hamiltonian with qubit-state-dependent energy shifts. Additionally, we introduce effective Lindblad operators to model the conditional states of the qubit atoms and the corresponding probabilities by postselecting state $|g\rangle$ of the auxiliary atom.} 

We assume that the dissipation of the system is described by  {the} Lindblad operators: ${L_{c_l}} = \sqrt {\kappa} {c_l}$, with $l=1, 2, 3$    {representing} the   {photon loss} of   {the} cavity modes with   {the same} dissipation rate $\kappa$; ${L_f} = \sqrt {{\gamma _f}} \left| f \right\rangle \left\langle {{E_1}} \right|$  and  ${L_g} = \sqrt {{\gamma_g}} \left| g \right\rangle \left\langle {{E_2}} \right|$ describe the decay of the auxiliary atom  with  rates $\gamma_f$ and $\gamma_g$, respectively; and ${L_k} = \sqrt {{\gamma}} \left| d \right\rangle \left\langle {{e}} \right|$ ($k=1,2$)   {describes} the decay of the  {qubit-encoding} atoms  with  rate  $\gamma$. We assume that the excited level  $\left |e \right \rangle$ decays to some level   {$\left | {d}  \right \rangle$, which, in fact, {may or may not be}} $\left | {1}  \right \rangle$ or $\left |{0} \right \rangle$, since the decay of either a cavity or an excited atom leads to a heralded error.
	
The standard master equation in the Lindblad form  for the composite system described by the Hamiltonian in Eq.~\eqref{HamiltonianT} can be given by~\cite{Gneiting2021Jump-time,Gneiting2022Unraveling}
\begin{align}\label{eq:rhoT}
\dot{\rho }_{T}\left ( t \right ) = &i\left [ \rho_T \left ( t \right ),H_T  \right ]
		+\frac{1}{2}\sum_{j} \left [2 L_j \rho_T \left ( t \right )L_j^\dagger\nonumber
		\right. \\ & \left. -\rho_T \left ( t \right )L_j^\dagger L_j -L_j^\dagger L_j\rho_T \left ( t \right )\right ],
\end{align}
where $\rho _{T} \left ( t \right ) $ represents the density matrix of the   {total} system. 
{Alternatively, the standard Lindblad
master equation can be recast in the form with the non-Hermitian Hamiltonian $H_{\rm NH}^T=H_T- \frac{i}{2}\sum_j {L_j^\dagger {L_j}}$ and the quantum-jump
terms $\sum_j L_j \rho_T(t)L_j^\dagger$, as it is done in quantum-trajectory approaches~\cite{Dalibard1992Wave-function,Carmichael1993Quantum,Molmer1993Monte}, as follows:
\begin{align}\label{eq:QTA}
\!\!\!\!\!\dot{\rho }_{T}\left ( t \right ) =& \mathcal{L}\dot{\rho }_{T}\nonumber\\
=&-i\left[H_{\rm NH}^T\rho_T \left ( t \right )-\rho_T \left ( t \right )H^{T\dagger}_{\rm NH}\right] \nonumber\\
&+\sum_{j} L_j \rho_T \left ( t \right )L_j^\dagger,
\end{align}
which can be used to study the effect of quantum jumps in relation to quantum exceptional points~\cite{Minganti2019Quantum} and to analyze the postselection on the number of quantum jumps 
within the hybrid-Liouvillian formalism~\cite{Minganti2020Hybrid}.}

For a weak classical driving field, i.e., {$\left \{ {\Omega/\Delta _{E_2}}, {\Omega/g}  \right \}\ll1$}, the   {excitations} of   {the} cavity modes and the excited states   {of the atoms} can be adiabatically  {eliminated, when} the system is initially prepared in the   {ground-state subspace.}
   Therefore, the ground-state evolution of the composite system can be described by an effective master equation as follows~\cite{kastoryano2011dissipative,reiter2012effective}:
	\begin{align}\label{eq:rho_eff}
		\dot \rho  =& i\left[ \rho ,H_{\rm eff} \right]
		+ \frac{1}{2} \sum\limits_j \left \{   2 L_{\rm eff}^j\rho {\left( L_{\rm eff}^j \right)}^\dagger\nonumber
		\right. \\ & \left.-\left [ {\left( {L_{\rm{eff}}^j} \right)}^\dagger L_{\rm{eff}}^j\rho  +\rho {\left( {L_{\rm{eff}}^j} \right)}^\dagger L_{\rm{eff}}^j\right ] \right \}.
	\end{align}
 Here $\rho$ denotes the ground-space density matrix of the composite  {system; $H_{\rm{eff}}$}  represents an effective Hamiltonian given by
	\begin{equation}\label{eq:Heff}
		{H_{{\rm{eff}}}} =  - \frac{1}{2}{V^\dag }\left[ {H_{\rm{NH}}^{ - 1} + {{\left( {H_{\rm{NH}}^{ - 1}} \right)}^\dag }} \right]V,
	\end{equation}
	and  $L_{\rm{eff}}^j$  are the effective Lindblad operators with
	\begin{equation}\label{eq:L_eff}
		L_{\rm{eff}}^j = {L_j}H_{{\rm{NH}}}^{ - 1}V,
	\end{equation}
while the non-Hermitian Hamiltonian $H_{\rm{NH}}$ governing the dynamics of the decaying excited states~\cite{reiter2012effective,Arkhipov2024Restoring}   { {can be given, in the quantum jump formalism, as}}
	\begin{eqnarray}\label{eq:H_NH}
		\!\!\!  H_{\rm{NH}} &= & {H_e} - \frac{i}{2}\sum\limits_j {L_j^\dagger {L_j}}\;\;\;\;\;\;\;\;\;\;\nonumber\\
		&=&\!\sum_{k=1,2} \!\Big[{\frac{\bar\Delta _e}{2} {\left|{e}\right\rangle _k}\!\left\langle{e}\right|} + \frac{g}{2}( {{c_1}+{c_2}+\sqrt{2}S_k {c_3}} ){{\left|{e}\right\rangle }_k}\!\left\langle{1}\right| \nonumber\\
		&&+{\rm H.c.}\Big]
+{{\bar \Delta }_{E_1}}\left| {{E_1}} \right\rangle \left\langle {{E_1}} \right| + {{\bar \Delta }_{E_2}}\left| {{E_2}} \right\rangle \left\langle {{E_2}} \right|\nonumber\\
		&& - \frac{{i\kappa }}{2} c_1^\dagger {c_1}
		+\frac{g_f}{\sqrt 2 }\left[ {\left( {{c_2} - {c_1}}\right)\left| {{E_1}} \right\rangle \left\langle f \right| + {\rm{H.c.}}} \right] \nonumber\\
		&&+ \sum_{l=2,3}{{\bar J}_l c_l^\dagger c_l}+ \frac{\Omega _{m}}{2}\left( {\left|{E_1}\right\rangle \left\langle{E_2}\right| + {{\rm H.c.}}} \right).
	\end{eqnarray}
Here,  the auxiliary parameters are as follows:
\begin{align}
{\bar \Delta _{E_1}} &= {\Delta _{E_1}} - i{\gamma _f}/2, \nonumber\\
{\bar \Delta _{E_2}} &= {\Delta _{E_2}} - i{\gamma _g}/2, \nonumber\\
{\bar \Delta _e} &= {\Delta _e} - i\gamma /2, \nonumber\\
{\bar J_{2}} &=  2\sqrt 2 J - i\kappa /2, \nonumber\\
{\bar J_{3}} &=  \sqrt 2 J - i\kappa /2.
\end{align}
	
	To achieve the nonlocal heralded gate, the composite system is confined within the zero- and single-excitation   {subspaces}. The effective Hamiltonian $H_{\rm eff}$ and the effective Lindblad operators $L_{\rm eff}^j$
	can be directly derived from Eqs.~\eqref{eq:Heff}--\eqref{eq:H_NH}. Specifically, $H_{\rm eff}$   {is given} as follows:
	\begin{equation}\label{eq:H_effN}
		{H_{\rm eff}} = \left| g \right\rangle \left\langle g \right| \otimes \sum\limits_{N = 0}^2 {{\Delta _N}} {\mathcal{P} _N},
	\end{equation}
	where $\mathcal{P} _N$ is a projection operator that projects   {the} two  {qubit-encoding} atoms   {onto} a state  with $N$ qubits in    {$\left | 1  \right \rangle$, while} $\Delta_N$ represents the $N$-dependent ac Stark shift,  {which} can be   {expressed} as
	\begin{align}\label{eq:DeltaN}
		{\Delta _N} =&  - \frac{\Omega ^2}{\gamma }\text{Re} \left \{ \frac{1}{\mathcal{X}_N }\left [ C{\widetilde \Delta }_e\left( {m + n} \right)\left( {{S_1} + {\widetilde J}_2{S_2}} \right)
		 \nonumber
		\right.\right. \\ & \left. \left.- 2{{\widetilde \Delta }_e}^2{{\widetilde J}_2}{S_1} - 2mn{C^2}{S_2}
		 \right ]  \right \},
	\end{align}
	where {$\rm Re$} denotes the real part of   {an} argument, and
$m\left ( n \right ) \in\{0,1\}$ denotes the number of  {the qubit-encoding} atoms in state $\left | 1  \right \rangle$   {and} coupled to cavity A~(C).
	  {Moreover,} the auxiliary parameters are as follows: 
	\begin{align}
C=&g^{2} /( \gamma \kappa  ),  \nonumber\\
C_f=&g_f^{2} /( \gamma \kappa  ),\nonumber\\
	\widetilde{\Omega}_{m}=&\Omega _{m}/\gamma,\nonumber\\
	{\widetilde J_{1}}=& 2\sqrt 2 J/\kappa - i /2,\nonumber\\
	{\widetilde J_{2}}=&\sqrt 2 J/\kappa - i /2,\nonumber\\
	\widetilde{\Delta }_{e}=& \Delta_{e}/\gamma -i/2,\nonumber\\
	\widetilde{\Delta }_{E_1}=&\Delta_{E_1}/\gamma -i\gamma_f/\left(2 \gamma  \right),\nonumber\\
	\widetilde{\Delta }_{E_2}=&\Delta_{E_2}/\gamma -i\gamma_g/\left( 2 \gamma  \right),\nonumber\\
	{S_1} =& {C_f}\left( {2i{{\widetilde J}_1} + 1} \right) - 2{\widetilde \Delta _{E_1}}{\widetilde J_1},\nonumber\\
	{S_2} =& 4i{C_f} - {\widetilde \Delta _{E_1}}( {2i{{\widetilde J}_1} + 1} ),\nonumber\\
Z = &4{\widetilde \Delta _{E_1}}{\widetilde \Delta _{E_2}} - \widetilde \Omega _{m}^2,\nonumber\\
\mathcal{X}_N  =& {{C_f}{{\widetilde \Delta }_{E_2}}{R_2} - {R_1}Z},\nonumber\\
		{R_1} =& {\widetilde \Delta _e}C\left( {m + n} \right)\left( {{{\widetilde J}_2} + 2{{\widetilde J}_1} + 2i{{\widetilde J}_1}{{\widetilde J}_2}} \right) \nonumber\\
		&- 2{C^2}mn\left( {2i{{\widetilde J}_1} + 1} \right)   - 4{\widetilde \Delta _e}^2{\widetilde J_1}{\widetilde J_2},\nonumber\\
    	{R_2} =& 4{\widetilde \Delta _e}C\left( {m + n} \right)\left[ {2i( {{{\widetilde J}_1} + 2{{\widetilde J}_2}}) + 1} \right] \nonumber\\
    	&- 32i{C^2}mn - 8{\widetilde \Delta _e}^2{\widetilde J_2}\left( {2i{{\widetilde J}_1} + 1} \right).
    \end{align}
 {The} effective Lindblad operators   {are expressed} as follows:
	\begin{align} \label{L-oper}
		L_{\rm eff}^g &= \left| g \right\rangle \left\langle g \right| \otimes \sum\limits_{N = 0}^2 {{r_{g,\scriptsize N}}{\mathcal{P} _{\scriptsize N}}},\nonumber\\
		L_{\rm eff}^f &= \left| f \right\rangle \left\langle g \right| \otimes \sum\limits_{N = 0}^2 {{r_{f,{\scriptsize N}}}{\mathcal{P} _{\scriptsize N}}},\nonumber\\
		L_{\rm eff}^{{c_l}} &= \left| f \right\rangle \left\langle g \right| \otimes \sum\limits_{N = 0}^2 {{r_{{c_l},{\scriptsize N}}}{\mathcal{P} _{\scriptsize N}}},\nonumber\\
		L_{\rm eff}^k &= \left| f \right\rangle \left\langle g \right| \otimes \sum\limits_{N = 1}^2 {{r_{k,{\scriptsize N}}}{{\left| d \right\rangle }_k}\left\langle 1 \right|{\mathcal{P} _{\scriptsize N}}},	
	\end{align}
	where $k=1$ ($k=2$)   {labels} the  {qubit-encoding} atom coupled to cavity A (C) in  state $|1\rangle$.
	 The corresponding effective decay rates $r_ {g,\scalebox{0.8}{\textit{N}}}$, $r_ {f,\scalebox{0.8}{\textit{N}}}$, $ r_{c_l,\scalebox{0.8}{\textit{N}}}$, and $r_ {k,\scalebox{0.8}{\textit{N}}}$  {are   {given by}}
		\begin{align}\label{eq:decay1}
		    {r_{g,N}} =& \frac{2\Omega \sqrt {{\gamma _g}} }{\gamma  \mathcal{X}_N  }\left[ C{{\widetilde \Delta }_e}\left( {m + n} \right)\left( {S_1} + {{\widetilde J}_2}{S_2} \right) \nonumber
			 \right. \\ & \left.- 2{{\widetilde \Delta }_e}^2{{\widetilde J}_2}{S_1} - 2mn{C^2}{S_2} \right],\nonumber\\
			 {r_{f,N}} = &{\Omega {{\widetilde \Omega }_m}{R_1}\sqrt {\gamma _f} }/{\gamma  \mathcal{X}_N  },\nonumber\\
			{r_{{c_1},N}} =& {2\sqrt 2 i}\delta \left[ {{\widetilde \Delta }_e}C\left( {{\widetilde J}_1} + {{\widetilde J}_2} \right)\left( {m + n} \right) \nonumber
			\right. \\ & \left.- 2{{\widetilde \Delta }_e}^2{{\widetilde J}_1}{{\widetilde J}_2} - 2{C^2}mn \right], \nonumber\\
			{r_{{c_2},N}} =& {\sqrt 2 }{\delta}\left[ 2{{\widetilde \Delta }_e}^2{{\widetilde J}_2} + 4i{C^2}mn \nonumber
			\right. \\ & \left.- C{{\widetilde \Delta }_e}\left( {1 + 2i{{\widetilde J}_2}} \right)\left( {m + n} \right) \right],\nonumber\\
			{r_{{c_3},N}} = & {C }{\delta}\left[ {{\widetilde \Delta }_e}\left( {1 - 2i{{\widetilde J}_1}} \right)\left( {m - n} \right) \right],\nonumber\\
			{r_{1,N}} = &\sqrt {2C}{\delta}[(1- 2 i {\widetilde J}_1)(nC-{{\widetilde \Delta }_e}{{\widetilde J}_2})],\nonumber\\
			{r_{2,N}} = & \sqrt {2C}{\delta}[(1- 2 i {\widetilde J}_1)(mC-{{\widetilde \Delta }_e}{{\widetilde J}_2})],\nonumber\\
            \delta=&{\sqrt {{C_f}} \Omega {{\widetilde \Omega }_m}}/({\sqrt \gamma  \mathcal{X}_N }).
		\end{align}

	For a weak field, driving the transition {$|E_2\rangle\rightarrow |E_1\rangle$} with ${\Omega _{\rm m}}/{\Delta _{E_2}} \ll1$, the ac Stark shift $\Delta _N$ and the effective decay rates  $r_ {i,{\tiny N}}$,   {shown in  Eqs.~\eqref{eq:DeltaN} and \eqref{eq:decay1},  {can be simplified:}
		\begin{align}\label{eq:decay2}
			{\Delta _N} =&  - \frac{\Omega ^2}{4{\Delta _{E_2}}} - \frac{{\widetilde \Omega }^2}{4\gamma }\text{Re} \left( \frac{Q}{C_f R + {\widetilde \Delta} _{E_1}Q} \right),\nonumber\\
			r_{f,N} =&  - \frac{\widetilde {\Omega}\vspace{1ex} {Q} \sqrt {{\gamma _f}} }{2 \gamma (C_f R + {\widetilde \Delta} _{E_1}Q)},\nonumber\\
			r_{g,N} =& \frac{\Omega \sqrt {\gamma _g} }{2 \Delta _{E_2}} + \frac{\widetilde {\Omega} \vspace{1ex} {Q} \sqrt {{\widetilde \gamma }_g}}{2\gamma  (C_f R + {\widetilde \Delta} _{E_1}Q)},\nonumber\\
			r_{{c_1},N} =& { 2 \sqrt {2} \delta' }\left[ 2{{\widetilde \Delta }_e}^2{{\widetilde J}_1}{{\widetilde J}_2} + 2{C^2}mn \nonumber
			\right. \\ & \left.	- C{{\widetilde \Delta }_e}\left( {{{\widetilde J}_1} + {{\widetilde J}_2}} \right)\left( {m + n} \right) \right],\nonumber\\
			{r_{{c_2},N}} =& {\sqrt {2} \delta' }\left[ 2i{{\widetilde \Delta }_e}^2{{\widetilde J}_2} - 4{C^2}mn \nonumber
			\right. \\ & \left.+ C{{\widetilde \Delta }_e}\left( {2{{\widetilde J}_2} - i} \right)\left( {m + n} \right) \right],\nonumber\\
			{r_{{c_3},N}} =& { \delta' }\left[ C{{\widetilde \Delta }_e}\left( {i + 2{{\widetilde J}_1}} \right)\left( {m - n} \right) \right],\nonumber\\
			{r_{1,N}} =&  \alpha' \sqrt {2C}{\delta}[(1- 2 i {\widetilde J}_1)(nC-{{\widetilde \Delta }_e}{{\widetilde J}_2})],\nonumber\\
			{r_{2,N}} =&  \alpha' \sqrt {2C}{\delta}[(1- 2 i {\widetilde J}_1)(mC-{{\widetilde \Delta }_e}{{\widetilde J}_2})],\nonumber\\
            \alpha'=& {i \widetilde \Omega \mathcal{X}_N}/[2 \Omega \Omega _{m}  (C_f R + {\widetilde \Delta} _{E_1}Q)],\nonumber\\
                        \delta'=&{ \widetilde \Omega \sqrt { C_f} }/[2\sqrt {\gamma }(C_f R + {\widetilde \Delta} _{E_1}Q)],\nonumber\\
		R =& 2{{\widetilde \Delta }_e}^2\left( { - i + 2{{\widetilde J}_1}} \right){{\widetilde J}_2}+8{C^2}mn \nonumber\\
		&- C{{\widetilde \Delta }_e}\left( { - i + 2{{\widetilde J}_1} + 4{{\widetilde J}_2}} \right)\left( {m + n} \right),\nonumber\\
		Q =& 4i{\widetilde \Delta _e}{\widetilde J_1}{\widetilde J_2}+ 2{C^2}\left( {i - 2{{\widetilde J}_1}} \right)mn \nonumber\\
		&+C{\widetilde \Delta _e}\left[ 2{{\widetilde J}_1}{{\widetilde J}_2} - i\left( {{{\widetilde J}_2} + 2{{\widetilde J}_1}} \right) \right]\left( m + n \right).
	\end{align}
We note that  {$\widetilde{\Omega } =\Omega \Omega _{m}/{(2\Delta _{E_2}) }$} is the effective Rabi frequency of the transition $|g\rangle\rightarrow|E_1\rangle$  and $\widetilde{\gamma_g }=\gamma_g\Omega _{m}^2/\left (2\Delta_{E_2}   \right )^2 $  is an effective decay rate of the excited state $|E_1\rangle$ to  $|g\rangle$.

In practice, the  auxiliary and the  {qubit-encoding} atoms can be different. Their atom-cavity cooperativities and decay rates  can be   parameterized by $C_f=\alpha C$ and $\gamma _f=\beta \gamma$. For
simplicity, we set $\alpha=\beta=1$ in all our numerical simulations to show the influence of the cooperativity $C$ on the system evolution.   {In this case,} $\Delta_N$ and  $r_{g,N}$  can be further simplified  {as:}
\begin{align}\label{eq:rg_fin}
	{\Delta _N} =&  - \frac{{\widetilde \Omega }^2}{4\gamma }\text{Re} \left( \frac{Q}{C_f R + {\widetilde \Delta} _{E_1}Q} \right),\nonumber\\
	{r_{g,N}} =& \frac{\Omega \sqrt {\gamma _g}}{2\Delta _{E_2}},
\end{align}
where the first term, $-\Omega^2/\left(4 \Delta_{E_2} \right) $, of $\Delta _N$ in Eq.~\eqref{eq:decay2} has  been removed,  {because} it
is independent of the state of the qubits  {and, thus, has} no influence on the phase gates. Furthermore,  the second term of $r_{g,N}$ has also been removed for $\widetilde{\gamma }_g \ll 1$,  {because} the decay of the  {auxiliary-atom} excited state to $\left | g  \right \rangle$ is suppressed by  {the} large detuning $\Delta_{E_2}$.

Each Lindblad operator shown in Eq.~\eqref{L-oper}, except $L_{\rm eff}^g$  {(i.e., the dephasing of $|g\rangle$), represents}    { various effective\textit{ dissipative processes}}, leading to the transition $\left | g  \right \rangle \to \left | f  \right \rangle$. These
are the dominant error factors that   {drive} the system out of its effective subspace. Fortunately,   {the} errors introduced by these dissipative processes can be inherently detected,  {because} the success of each nonlocal two-qubit gate is heralded by   {the measurement result $\left | g  \right \rangle$ of} the auxiliary atom.
For heralded gates, these detectable decays have no effect on the fidelity, but decrease their success probability.

The \textit{success probability} $P$ of detecting the auxiliary atom in  state $\left | g  \right \rangle$ can be obtained by solving the effective Lindblad master equation,   {given} in Eq.~\eqref{eq:rho_eff}, with the following definition
\begin{equation}\label{eq:P}
		P = \sum\limits_{N = 0}^2 {{\rm Tr}\left[ {\left( {\left| g \right\rangle \left\langle g \right| \otimes {\mathcal{P} _N}} \right)\rho \left( t \right)} \right]},
\end{equation}
where ${\rm Tr}$ is the trace operation over the subspace spanned by the ground states of the auxiliary and  {qubit-encoding} atoms.

After the measurement on the auxiliary atom, the \textit{conditional} density operator of the two  {qubit-encoding} atoms is reduced to
\begin{eqnarray}\label{eq:rho_qubit}
	\rho _{\rm qubit}{\left( t \right)} =&& \frac{1}{P}\sum\limits_{N,N' = 0}^2 {e^{ - i({\Delta _N} - {\Delta _{N'}})t}}{e^{ - ({\Gamma _N} + {\Gamma _{N'}})t/2}} \nonumber\\
	&&\times {\mathcal{P} _N}\left[ \left\langle g \right|\rho \left( 0 \right)\left| g \right\rangle  \right]{\mathcal{P} _{N'}}.
\end{eqnarray}
Here the \textit{total decay rate} $\Gamma _N$ for $N$  {qubit-encoding} atoms in  state $|1\rangle$   {is found to be}
\begin{eqnarray}
 {\Gamma _N} = {\left| {{r_{f,N}}} \right|^2} + \sum\limits_{l = 1}^3 {{{\left| {{r_{{c_l},N}}} \right|}^2}}  + m{\left| {{r_{1,N}}} \right|^2} + n{\left| {{r_{2,N}}} \right|^2},\;\;\;\;\;
\end{eqnarray}
where $r_ {g,{\tiny N}}$,  $r_ {f,{\scriptsize N}}$,  $ r_{c_l,{\scriptsize N}}$, and $r_ {k,{\scriptsize N}}$ are   {the} effective decay rates   {given} in Eq.~\eqref{eq:decay1}.
\textit{By properly controlling the evolution time and measuring the auxiliary atom, we can in principle achieve  {a  two-qubit} nonlocal  CPHASE gate in a heralded way}, as described below. The success probability of the gate is   {equal} to that   {of} projecting  the auxiliary  {atom}   {onto} state $|g\rangle$.  All basic
symbols used in this paper are shown in Table \ref{tab1}.

\bigskip	
\section{Heralded nonlocal CPHASE gate and its performance}\label{se:simulation}
The effective Hamiltonian in Eq.~\eqref{eq:H_effN} shows that the energy   {shift} depends on the number of  { qubit-encoding} atoms in    state $|1\rangle$ when the auxiliary  {atom} is in   {the} state $|g\rangle$. Therefore, the time evolution
under this effective Hamiltonian gives rise to different dynamical phases
for   {the} two qubits in   {the} states   {$|00\rangle$, $|10\rangle$, $|01\rangle$, and $|11\rangle$.} By choosing a suitable evolution time and   {then} performing single-qubit transformations, we can achieve a phase flip of the qubit state $\left | 11  \right \rangle$, while leaving the other three states unchanged, which   {achieves} the \textit{heralded nonlocal} CPHASE gate on  {the} two spatially separated {atom qubits}.

The detrimental effect of \textit{dissipative processes} on the CPHASE  gate, represented by the state flip of the auxiliary  {atom, can} be inherently removed by projecting   {the auxiliary  {atom} onto the} state $|g\rangle$, while the state-dependent decay   {rate} $\Gamma_N$ of the  {qubit-encoding} atoms and   {the} finite spontaneous decay rate  ${\widetilde \gamma}_g > 0$   {can} introduce extra errors. Therefore, we can improve the gate fidelity by modifying the system to achieve a state-independent decay rate, i.e., $\Gamma_0=\Gamma_1=\Gamma_2$. The state-independent total decay rate $\Gamma_N$, in the limit  {$\left\{ {G,C} \right\} \gg 1$, {where}}  $G=J/\kappa$,  can be given by
\begin{equation}
	\Gamma _N = \Gamma  = \frac{{\widetilde \Omega }^2}{2\gamma }\frac{1}{\alpha C},
\end{equation}
 {where} the detunings are changed to
\begin{eqnarray}
	\frac{\Delta _{E_1}}{\gamma } &=& {\alpha  C D/\sqrt{2}} ,\label{eq:D_E1} \nonumber\\
	\frac{{{\Delta _e}}}{\gamma } &=& \frac{ - 2 + C\left(  {\bar{G}^2} - 4{D\bar{G}} \right)}{2\sqrt 2 \left( {\bar{G}-2 D } \right)},\label{eq:D_e}
\end{eqnarray}
where  {$\bar{G} =1/G$ and} $D=\sqrt {\beta/\alpha C}$   {are two auxiliary parameters}.
The corresponding energy   {shift}  {can be rewritten as:}
	\begin{align}\label{eq:Delta_012}
		{\Delta _0} =& -\Gamma \frac{\left( 4D-\bar{G} \right)}{8\sqrt 2 }, \nonumber\\
		{\Delta _1} =& -\frac{\Gamma }{\sqrt 2 }\frac{2D-\bar{G}}{2/C + \bar{G}^2 - D\bar{G} + 2{D^2}},\nonumber \\
		{\Delta _2} =&  -\frac{\Gamma }{\sqrt 2 }\frac{2D-\bar{G} }{1/C +  {\bar{G}^2/2} - D\bar{G} + 2D^2},
	\end{align}
%		{\Delta _2} =&  -{\Gamma }\frac{\sqrt{2} \left(  2D-1/G \right) }{2/C + 1/G^2 - 2 D/G + 4D^2},
where $\Delta_0$ approaches \textit{zero} for $\left\{ {G,C} \right\} \gg 1$, while $\Delta _1$ and $\Delta _2$ are nonzero and approximately equal  {to each other.} This property can be used to achieve a heralded nonlocal CPHASE gate by a driving pulse with duration 
\begin{align}
T_\pi =\frac{\pi}{|\Delta _2|}.
\end{align}

In practice, we can further decrease the \textit{gate error} to arbitrarily small by performing unitary single-qubit rotations on each  {qubit-encoding} atom, which depends on the dynamical evolution of the composite system. The duration of the driving pulse length is chosen to be
	\begin{align}\label{eq:t_CZ}
		t_{\text{CZ}} = \dfrac{\pi}{\left| {{\Delta _2} - 2{\Delta _1} + {\Delta _0}} \right|} ,
	\end{align}
	and the single-qubit  rotation  on each qubit after  {applying} the pulse  {reads:}
	\begin{eqnarray}\label{eq:U}
\mathcal{U} \left | 0  \right \rangle &=&\exp({i\Delta _0 t_{\text{CZ}}/2})\left | 0  \right \rangle,\nonumber\\
\mathcal{U} \left | 1  \right \rangle &=&\exp[{i\left ( 2\Delta _1-\Delta _0 \right )  t_{\text{CZ}}/2}]\left | 1  \right \rangle.
	\end{eqnarray}
These processes result in a \textit{phase flip}
of the state $|11\rangle$,
while leaving the other three states   {(i.e., $|00\rangle$, $|10\rangle$, and $|01\rangle$)} unchanged.

 \begin{figure}[tbp]
	\centering
\includegraphics[width = 0.48\textwidth]{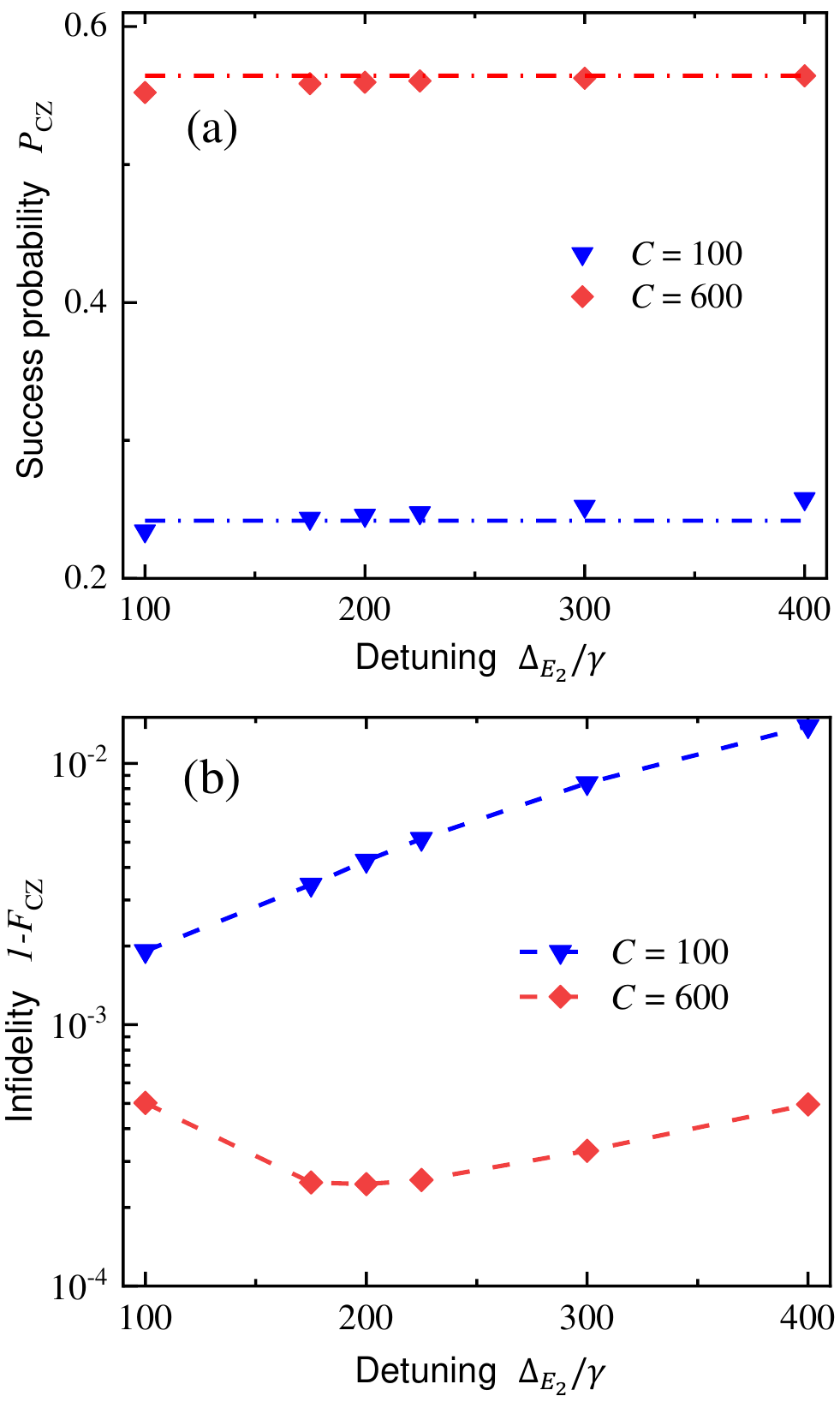}
	\caption{Numerical simulations for the success probability and infidelity of the heralded nonlocal CPHASE gate with two cooperativities $C=100$~(blue down-triangles) and $C=600$~(red diamonds). (a) The success probability, $P_{\rm CZ}$, of the gate as a function of  {the}   detuning $\Delta_{E_2}$. Simultaneously, we also plot the analytical success probability (curves), which
	 {is in good agreement} with the numerical values. (b)  Infidelity, $(1-F_{\rm CZ})$ of the CPHASE gate vs   {the}  detuning $\Delta_{E_2}$. In both panels,   {we have set:} $\lambda=10$,  $\gamma_g=\gamma_f=\gamma=0.1\kappa$, $g=g_f$, $C=g^2/(\kappa \gamma)$, $\lambda=J/(\kappa \sqrt{C})$, $\alpha=\beta=1$, $\Omega=\Delta_{E_2}/(6C^{1/4})$, and $\Omega_m=4\gamma C^{1/4}$.}
	\label{fig:3cav_simulation}
\end{figure}

The \textit{success probability of the heralded nonlocal} CPHASE gate equals that   {of finding} the auxiliary  {atom} in   {the} state $|g\rangle$   {at the end of the gate {operation, and} can be given by}
	\begin{equation}\label{eq:PCZ0}
		{P_{\text{CZ}} = \exp \left( { - \Gamma {t_{\text{CZ}}}} \right).}
	\end{equation}
      {It} can be   {further} approximated as
	\begin{equation}\label{eq:PCZ1}
		{P_{\text{CZ}}} = 1 - {Z_p}\frac{\pi }{{\sqrt C }},
	\end{equation}
    for $\left \{ C,G \right \}\gg 1$, where the scaling factor $Z_p$, with
    $ \lambda = G/\sqrt{C}$ and  $d = \sqrt{\beta / \alpha } $, can be   {given} as
    \begin{equation}\label{eq:Z_p}
    	{Z_p} = \sqrt 2 d + \frac{{\left( {1 + 2{\lambda ^2}} \right)}^2}{\sqrt 2 d{\lambda ^2}{\left( {1 - 2d\lambda } \right)}^2} + \frac{3 + 6{\lambda ^2}}{\sqrt 2 \lambda\left( { { 2d\lambda-1} } \right)}.
    \end{equation}
As long as $\lambda \gg 1 $, the success probability $P_{\text{CZ}}$ remains almost constant for a given $C$. In   {fact}, we need to select  appropriate parameters to ensure that the success probability of the heralded nonlocal CPHASE gate remains relatively high, while its error is arbitrarily small.

To demonstrate the feasibility of our protocol, we perform numerical simulations of the evolution of the composite system with the full master equation in Eq.~\eqref{eq:rhoT}, instead of the effective master equation in Eq.~\eqref{eq:rho_eff}. The initial state of our composite system is  {assumed to be}
	\begin{equation}
		{\left| \Psi  \right\rangle _{\rm ini}} = {\left| \Phi  \right\rangle _{\rm ini}} \otimes \left| {\rm vac} \right\rangle,
	\end{equation}
	where $\left| \Phi  \right\rangle _{\rm ini}$ represents the initial state of the auxiliary and  {qubit-encoding} atoms,   {given by}
	\begin{equation}
		 {{\left| \Phi  \right\rangle _{\rm ini}} = \left| g \right\rangle \left[ {\prod\limits_{k = 1}^2 \left| + \right\rangle_{k} } \right]},
	\end{equation}
{where $\left| + \right\rangle_{k}={{{\left( {\left| 0 \right\rangle_{k}  + \left| 1 \right\rangle_{k} } \right)}}/\sqrt{2}}$,} $\left | \rm vac \right \rangle $ is the vacuum state of the three coupled cavities. We solve the master equation with the  {QuTiP package}~\cite{johansson2012qutip,johansson2013qutip}, and calculate the \textit{success probability} {($P_{\rm CZ}$)} and \textit{fidelity} {($F_{\rm CZ}$)} of the gate with the following  expressions:
%	\begin{equation}\label{eq:PCZ_num}
%		{P_{\rm CZ}} = \sum_{N=0}^{2}  {{\rm Tr}\left[ {\left( {\left| g \right\rangle \left\langle g \right| \otimes \mathcal{P} _N \otimes \mathcal{I} } \right){\rho _T}\left( {{t_{CZ}}} \right)} \right]},
%	\end{equation}
%     \begin{equation}\label{eq:rho_qubit_cz}
%     	{\rho _{\rm qubit}}\left( {{t_{CZ}}} \right) = \frac{1}{P_{CZ}}{\rm Tr_{ cav}}\left[ {\left\langle g \right|{\rho _T}\left( t_{CZ} \right)\left| g \right\rangle } \right],
%     \end{equation}
%     \begin{equation}\label{F_CZ}
%     	{F_{\rm CZ}} = \left\langle \psi  \right|\left( {\mathcal{U} \otimes \mathcal{U}} \right){\rho _{\rm qubit}}\left( {{t_{CZ}}} \right){\left( {\mathcal{U} \otimes \mathcal{U}} \right)^\dag }\left| \psi  \right\rangle,
%     \end{equation}
	\begin{eqnarray}
		{P_{\rm CZ}} &=& \sum_{N=0}^{2}  {\rm Tr}\left[ {\left( {\left| g \right\rangle \left\langle g \right| \otimes \mathcal{P} _N \otimes \mathcal{I} } \right){\rho _T}\left( {{t_{CZ}}} \right)} \right], \;\;\;\;\; \\ \label{eq:PCZ_num}
     	{F_{\rm CZ}} &=& \left\langle \psi  \right|\left( {\mathcal{U} \otimes \mathcal{U}} \right){\rho _{\rm qubit}}\left( {{t_{CZ}}} \right){\left( {\mathcal{U} \otimes \mathcal{U}} \right)^\dag }\left| \psi  \right\rangle, \;\;\;\;\;\; \\ \label{F_CZ}
          	{\rho _{\rm qubit}}\left( {{t_{CZ}}} \right) &=& \frac{1}{P_{CZ}}{\rm Tr_{ cav}}\left[ {\left\langle g \right|{\rho _T}\left( t_{CZ} \right)\left| g \right\rangle } \right], \;\;\;  \label{eq:rho_qubit_cz}
     \end{eqnarray}
where ${\rm Tr}$  and ${\rm Tr_{cav}}$  are trace operations over the composite system and the cavities, respectively, and $\mathcal{I} $ is the identity operator   {for} the three cavities.

	The success probability $P_{\rm CZ}$ and  {the gate error (infidelity), $1-F_{\rm CZ}$, are} shown in Fig.~\ref{fig:3cav_simulation} as a function of the detuning $\Delta_{E_2}/\gamma$ for two different cooperativities $C=100$ and $C=600$. In  our numerical simulations, we set $\lambda = 10$ to   {reduce} the influence of   {the} off-resonant modes $c_2$ and $c_3$ on the gate error. Meanwhile, we assume that ${\gamma _g} = {\gamma _f}$, $\kappa  = 10\gamma $, $\alpha  = \beta  = 1$, $\Omega = \Delta _{E_2}/\left ( 6 C^{1/4 } \right ) $, and $\Omega_{m } = 4\gamma C^{1/4 }$.
	
	The detunings  $\Delta_{E_1}$ and  $\Delta_{e}$, given in Eq.~\eqref{eq:D_e}, are tuned to   {achieve} a   \textit{{total} qubit-independent decay rate}.
The numerical results~(marked by symbols)   {of} the success probability $P_{\rm CZ}$ are in agreement with the analytical ones determined by Eq.~\eqref{eq:PCZ1},   {as} shown in Fig.~\ref{fig:3cav_simulation}(a).
The success probability $P_{\rm CZ}$ is almost constant   {for a given cooperativity $C$} and
	gradually increases {with increasing} $C$. For the aforementioned parameters, $P_{\rm CZ}=0.56$ can be achieved for $C=600$.

The fidelity of the heralded nonlocal two-qubit gate, which is conditional on the detection of  {the} auxiliary  {atom in the} state $|g\rangle$, can approach unity   {in principle}. The finite length of the driving field in combination with the finite effective decay from  {$\left | E_2  \right \rangle$ to $\left | g  \right \rangle$}  {can} introduce    undetectable errors. 	
Theoretically, the former  {error} leads to    {a} nonadiabatic error of the gate,   {but which} can be suppressed by properly tuning the Rabi frequency $\Omega$ of the driving field. At the same time, the latter error can be decreased by increasing the detuning $\Delta_{E_2}$.  { For a cooperativity   } $C=100$, the gate error increases with the detuning $\Delta_{E_2}$, due to the {increase in} $\Omega$ and thus   {in the} nonadiabatic error, and can be less than $2\times10^{-3}$ for
$\Delta_{E_2}/\gamma=100$. For a larger cooperativity $C=600$, the gate error first decreases and then  {increases} with   { increasing  detuning} $\Delta_{E_2}/\gamma$.  {A gate error} below $3\times10^{-4}$ can be achieved for $C=600$ and $\Delta_{E_2}/\gamma=180$,   {as} shown in Fig.~\ref{fig:3cav_simulation}(b).

\begin{figure}[tbp]
\centering \includegraphics[width=8cm]{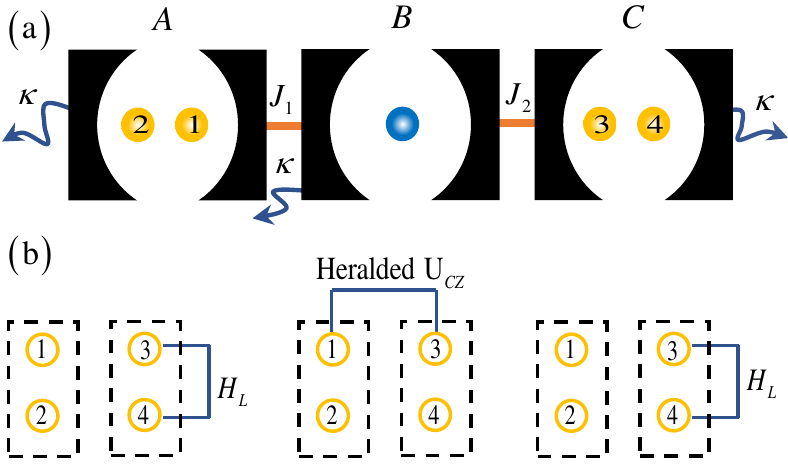}
\caption{(a) Schematic diagram of heralded nonlocal two-qubit quantum gates within a decoherence-free subspace. (b) Implementation  {scheme} of  {the CNOT gate}. $H_L$ represents    {the} Hadamard operation on a logical qubit consisting of  atoms  3 and 4,  and $U_{\rm CZ}$ represents the nonlocal CPHASE gate on  atoms 1 and 3 that couple to cavities A and C, respectively.}
		\label{fig3}
	\end{figure}
	
	\section{Heralded nonlocal quantum gates encoded in a decoherence-free subspace}
	\label{se:DFS}
In this section,  we  focus on the implementation of heralded single- and two-qubit gates on logical qubits that are robust against collective random dephasing errors, stemming from  {the}  fluctuations of the external fields  {and, thus, resulting in} uncontrolled energy shifts~\cite{lidar1998decoherence}.  In the case of collective dephasing,  the symmetry properties of the errors allow to identify a DFS in the Hilbert space of a  two-physical-qubit system~\cite{Wu2005Holonomic,You2005Correlation,Miccuda2020Decoherence-Resilient,Monz2009Realization,Qiao2022Generation}, where  {the} two logical
basis states can be $\left| 0_L  \right\rangle = \left| 01  \right\rangle$ and $\left| 1_L  \right\rangle=\left| 10  \right\rangle$, and
a memory-time enhancement of two orders of magnitude has been experimentally demonstrated for ion-trap systems~\cite{Monz2009Realization}.

Suppose that   {the}  {qubit-encoding} atoms 1 and 2 (3 and 4) are coupled  to  cavity A (C) and encode a logical qubit.   {Cavities} A and C interact with cavity B through two short fibers or superconducting coaxial cables,   {as} shown in Fig.~\ref{fig3}.   {We assume that there is an auxiliary atom coupled to cavity B.} The coupling rate  between cavities A~(C) and B is $J_1$~($J_2$), and all {three} cavities decay with   {the same} rate $\kappa$.

In principle, a CPHASE {gate, $U^{\text CZ}_L$, on}   {these} two logical qubits, given by {$U^\text{CZ}_L={\rm exp}{(i\pi\left| 1_L 1_L \right\rangle }{\left\langle 1_L 1_L \right| )}$} can be achieved with a  heralded nonlocal CPHASE gate $U^\text{CZ}_{1,3}$ on   {the} atom pair (1, 3) from two logical qubits.  The gate $U^\text{CZ}_{1,3}$ can be implemented with the same method described in   {the} previous {sections}, while the other two atoms   {need to} be decoupled from the cavities (i.e., by  modifying  their  detunings) during the  CPHASE gate operation. Furthermore, the controlled-NOT (CNOT) gate on two nonlocal logical qubits can be constructed by sandwiching the CPHASE gate with two Hadamard operations on the same logical qubit as follows:
\begin{equation}
		\text{CNOT}_L = \left( {I \otimes {H_L}} \right) \times \left(U^\text{CZ}_{13} \right) \times \left( {I \otimes {H_L}} \right),
	\end{equation}
where $I$ is the identity on the first logical qubit, $U^\text{CZ}_{13}$ is a nonlocal CPHASE gate performed on   {the} atom pair (1, 3),  and $H_L$ performs  {the} Hadamard transformation on the second logical qubit,   {as} shown in  Fig.~\ref{fig3}.

The  {operation of the} Hadamard gate on a logical qubit is nontrivial and changes the entanglement between two   {physical} atoms encoding a logical qubit. The logical Hadamard gate can be  implemented  by a two-qubit CNOT gate in combination with single-qubit rotations on two  {qubit-encoding} atoms as follows~\cite{Zwerger_2017}:
    \begin{align}
    	{H_L} =& \left[ {\left( {HSHZ} \right) \otimes \left( {HSH} \right)} \right] \text{CNOT}_{34}\nonumber\\
    	&\times \left[ {\left( {HSX} \right) \otimes X} \right],
    \end{align}
    where the    {gate} $S = {\rm diag}\left( {1,i} \right)$, in the computational basis $\{|0\rangle, |1\rangle\}$, denotes a rotation around the $z$-axis by an angle  {$ \pi /2 $; }$H$ is the standard Hadamard transformation on a single physical  {qubit; while} $X$ and $Z$ are Pauli operators.   {The  CNOT$_{34}$ gate,  with the control atom 3 and the target atom 4, can be implemented by}
	\begin{equation}
		\text{CNOT}_{34} = {H_4}U^\text{CZ}_{34}{H_4},
	\end{equation}
where $H_4$   {represents}  {the} Hadamard transform on 	the  qubit  { 4, and} $U^\text{CZ}_{34}$ is  {the} heralded CPHASE gate  {acting} on qubits 3 and 4 that  {are} coupled to the same cavity.

The heralded  CPHASE gate {$U_{34}^{\rm CZ}$}  {acting} on   qubits 3 and 4 can be achieved  {in a setup}   {similar} to  {that shown in} Fig.~\ref{fig1}, except   {that} cavity A   {is decoupled} from  cavity B, i.e., $J_1=0$ and $J_2=J$, and
the heralded nonlocal  CPHASE gate   {is modified} to   {become} a \textit{compact} one, as described in Ref.~\cite{qin2017heralded}.

In order to explicitly describe the dynamics of the  \textit{composite system} consisting of two cavities and three atoms, we perform a transformation
  {for} the two cavity modes and introduce   {the} symmetric and antisymmetric optical
modes, $a_{\pm}=\left(a_{B}\pm a_{C}\right)/\sqrt{2}$. The total
	Hamiltonian is $H_{T}=H_{e}+V+V^{\dag}$, where $V$ is the same as in Eq.~\eqref{eqHe1}, while $H_{e}$ is changed to
	\begin{align}\label{eq:He1}
		H_{e}=&\sum_{k=1}^{2}\left\{\Delta_{e}|e\rangle_{k}\langle e|+\frac{g}{\sqrt{2}}\left[\left(a_{+}-a_{-}\right)|e\rangle_{k}\langle 1|+\text{H.c.}\right]\right\}\nonumber\\
		&+\Delta_{E_1}|E_{1}\rangle\langle E_{1}|+\Delta_{E_2}|E_{2}\rangle\langle E_{2}|+2Ja_{+}^{\dag}a_{+}\nonumber\\
		&+\frac{g_{f}}{\sqrt{2}}\left[\left(a_{+}+a_{-}\right)|E_{1}\rangle\langle f |+\text{H.c.}\right]\nonumber\\
		&+\frac{\Omega_{m}}{2}\left(|E_{1}\rangle\langle
		E_{2}|+\text{H.c.}\right).
	\end{align}
For \textit{large detunings} ({${\Delta _{E_1}\gg\Omega }$} and ${\Delta _{E_2}\gg\Omega _m}$) and a \textit{large coupling strength} ($J\gg g_f$) between cavities B and C, we can adiabatically eliminate   {the} excited states $|E_1\rangle$ and $|E_2\rangle$ and   {then} obtain a three-photon resonant Raman transition from {$\left | {g}  \right \rangle$ to $\left | {f}  \right \rangle$}, by choosing a driving field with frequency
	\begin{align}\label{eq:omegaL2}
{{\omega _{L}} =  	{\omega _c}-{\omega _m}+{\omega _f} - {\omega _g}- J.}
	\end{align}
  {Such a \textit{three-photon resonant Raman transition}} is resonantly mediated by   {the} antisymmetric  mode $a_-$, while  detuned by  $2J$ from   {the} symmetric  mode $a_+$.

    	\begin{figure}[!ht]
    	\centering
\includegraphics[width = 0.48\textwidth]{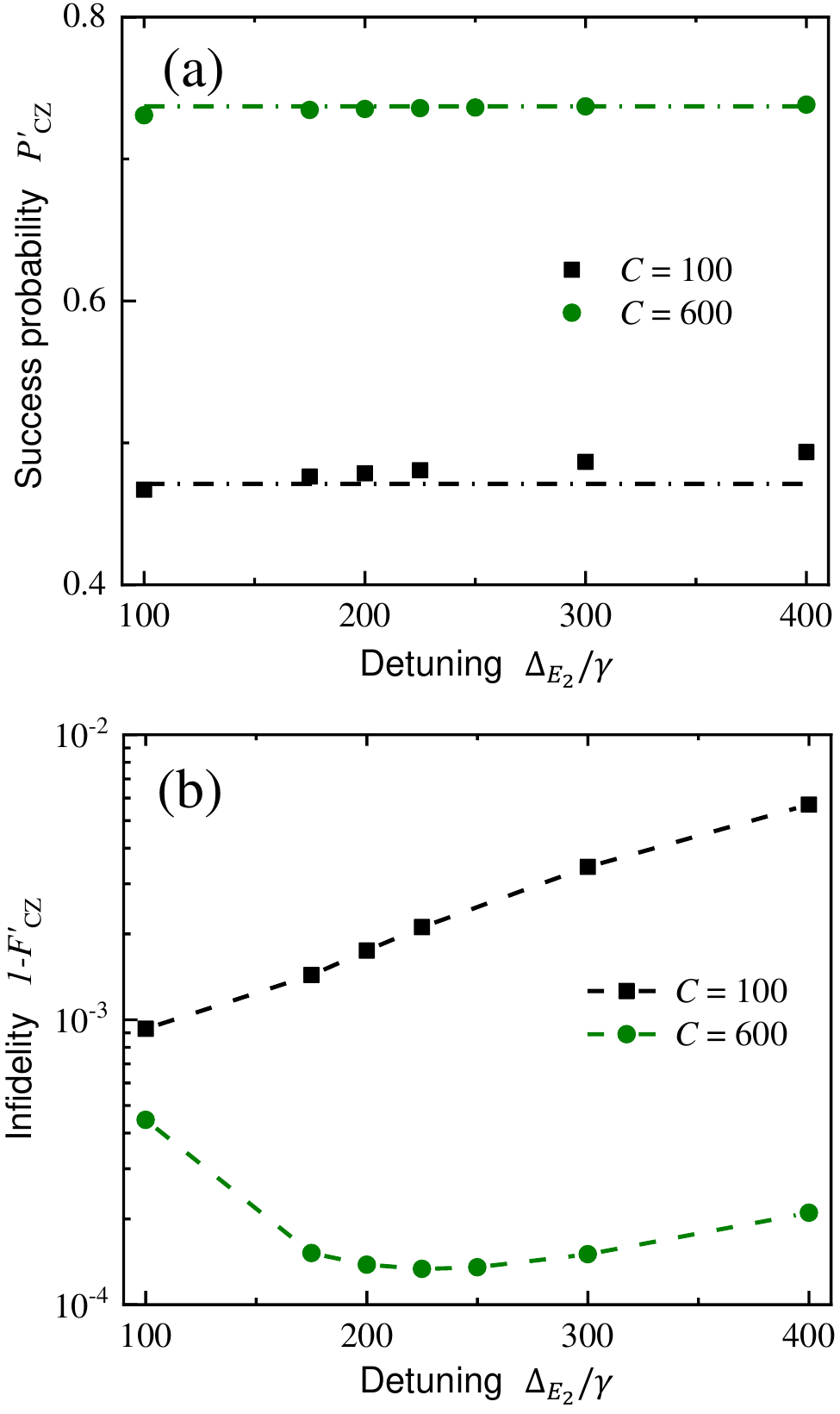}
    	\caption{Numerical simulations for the heralded CPHASE gate on two  {qubit-encoding} atoms,  the logical qubit,  with two  {cooperativities: $C=100$~(black squares)}  and $C=600$~(olive solid circles). (a) The success probability  $P'_{\rm CZ}$ as a function of  {the} detuning $\Delta_{E_2}/\gamma$. Simultaneously, we also plot the analytical results  {(shown by curves)}, which  {match well} with the numerical ones. (b)  {Infidelity $1-F'_{\rm CZ}$}  vs {the} detuning $\Delta_{E_2}/\gamma$.  All  {the} system parameters and the initial state are the same as  {those} assumed in Fig.~\ref{fig:3cav_simulation}, except $\lambda=1.84$.}
    	\label{fig4}
    \end{figure}

Following the procedure  in  Sec.~\ref{se:Hera_CZ}, we can implement the heralded near-deterministic CPHASE gate
on   {the}  {qubit-encoding} atoms 3 and 4   {in} the same cavity, which has been discussed in dissipative QED systems~\cite{qin2017heralded}. We can completely remove   {the} gate errors introduced by the qubit-dependent decay rate by modifying the detunings $\Delta _e$ and $\Delta_{E_1}$ to  {be:}
	\begin{align}
		\label{eq:Deltae1twocav}
		\frac{\Delta _e}{\gamma}&=\frac{1}{2\left(2D_1+\bar{G}\right)},\\
		\label{eq:DeltaE2twocav} \frac{\Delta _{E_1}}{\gamma}&=\alpha
		C\left(D_1+\bar{G} \right),
	\end{align}
	where $D_1=\sqrt{\left[\bar{G}^2+\beta/\left(\alpha
		C\right)\right]/2}$.  In the limit
	$\left\{ G,C \right\} \gg 1$,  the effective Hamiltonian   {driving} the evolution of the \textit{composite system}  can be described as
	\begin{equation}\label{eq:effH1}
		H_{\text{eff}}=|g \rangle\langle
		g|\otimes\sum_{n=0}^{2}\Delta'_{n}\mathcal{P}_{n},
	\end{equation}
	where $\mathcal{P}_{n}$   {is a projector onto} the states with $n$  qubit-encoding atoms in  state $|1\rangle$.
	The corresponding energy   {shift} $\Delta'_{n}$   {is given by}
	\begin{align}\label{eq:deltan_twocav}
		\Delta'_{0}=&-\frac{\Gamma D_1}{2},\\
		\Delta'_{n>0}=&-\frac{\widetilde{\Omega}^{2}}{2\gamma}\frac{n\left(2D_1+
			\bar{G}\right)}{\alpha C\left(4nD_1^2+2nD_1\bar{G}+1/C \right)},
	\end{align}
where $\Delta'_{0}$ approaches \textit{zero}, while $\Delta'_{1}\simeq\Delta'_{2}$ with {$|\Delta_{1}^{\prime}|\simeq|\Delta_{2}^{\prime}|\gg|\Delta_{0}^{\prime}|$} for $\left\{ G,C \right\} \gg 1$. Therefore, we can implement a CPHASE gate on  atoms 3 and 4 by properly
tuning the duration of the driving pulse  in combination with   {the} single-qubit rotations, according to Eqs.~(\ref{eq:t_CZ}) and (\ref{eq:U}), after replacing $\Delta_{n}$ with $\Delta'_{n}$.

The success probability $P'_{\rm CZ}$ and   {the} error ($1-F'_{\rm CZ}$) of the CPHASE gate on two  {qubit-encoding} atoms coupled to   {the same} cavity are of the same formalism as  {those} described in Eqs.~\eqref{eq:PCZ_num} and \eqref{F_CZ}, while the density matrix describes the composite system consisting of  { three atoms} and two cavities.

We numerically calculate  $P'_{\rm CZ}$ and ($1-F'_{\rm CZ}$) and   {demonstrate} their dependence on   {the} detuning $\Delta_{E_2}/\gamma$ for   {different} cooperativities ($C=100$ and $C=600$), shown in Fig.~\ref{fig4}. All  {the} system parameters and the initial state are the same as  {those} assumed in Fig.~\ref{fig:3cav_simulation}, except   {for} $\lambda=1.84$. {The success probability $P'_{\rm CZ}$ increases with  {increasing} $C$ and can be larger than that of the heralded nonlocal CPHASE gate with $P'_{\rm CZ}=0.74$ for $C=600$.} Meanwhile, the gate error decreases with $C$ and shows a dependence on detuning $\Delta_{E_2}/\gamma$, similar to that of the nonlocal CPHASE gate. For $C=600$, the gate error $1-F'_{\rm CZ}$ can be suppressed to $1.2\times10^{-4}$ for $\Delta_{E_2}/\gamma=220$. Therefore, the Hadamard gate in combination with the nonlocal CPHASE gate can be faithfully implemented with the cavity-coupled system in a heralded way.

%%%%%%%%%%%%%%%%%%%%%%%%%%%%%%%%%%%%%%%%%%%%%%%%%%%%%%%%%%%%%%%%%%%%%%%%%%%%%%%%%%%%%%%%%%%%%%%%%%%%%%%%%%%%%%%%%%%%%%%%%%%%%%%%%%%%%%%%%%%%%%%%%%%%%%%%%%%%%%%
	
\section{Discussion and summary}
\label{se:conclusions}
 {Our} protocol   {generalizes} the previous  {proposal of} heralded CPHASE gates~\cite{borregaard2015heralded,qin2017heralded} on qubits coupled to the same cavity to {a nonlocal case} by dynamically controlling the evolution rather than by scattering and measuring single photons. The integrated error detection eliminates the limitation    {of} single-photon sources and measurements~\cite{reiserer2015cavity},   and enables a high fidelity of the heralded CPHASE gates at the cost of a smaller success probability. Furthermore, we apply our heralded nonlocal CPHASE gate to heralded single- and two-qubit quantum gates within a DFS that is immune to collective dephasing   {noise}.
The heralded nonlocal CPHASE gates on qubits belonging to different cavities are suitable for interconnecting individual quantum processors for distributed quantum computing~\cite{Jiang2007Distributed} and quantum repeater networks~\cite{wehner2018quantum,yan2021survey}.

{Our  protocol} can be   {experimentally} implemented with neutral or artificial atoms coupled to  various cavities~\cite{buluta2011natural}. As an example, we consider ultracold $^{87}$Rb atoms coupled to optical cavities~\cite{borregaard2015heralded}. The relevant energy levels can be encoded   {as: the} two ground states
$\left | g \right \rangle$~($\left | 0 \right \rangle $) and $\left | f \right \rangle  $~($\left | 1 \right \rangle$),  {corresponding} to   {the} atomic levels $\left | F=1, m_f=1  \right \rangle$ and $\left | F=2, m_f=2  \right \rangle  $ of $5^2S_{1/2}$, respectively;  {and the} two excited states $\left | E_2 \right \rangle  $ and  $\left | E_1 \right \rangle  $($\left | e \right \rangle $),  {corresponding} to $\left | F=2, m_f=2  \right \rangle$ and $\left | F=3, m_f=3  \right \rangle$ of $5^2P_{3/2}$, respectively.    

{Optical cavities}  with high-$Q$ factors have
recently been widely used for quantum information technology~\cite{hacker2016photon,takahashi2020strong,kimble2008quantum}. The coupling strength $g$ between a cavity and an atom depends inversely on the cavity mode volume, i.e., $g\propto 1/\sqrt{V}$   {and}  {can, thus,} be significantly enhanced for small   {mode volume} cavities, such as fiber Fabry-Perot  cavities~\cite{brekenfeld2020quantum}, photonic crystal cavities~\cite{tiecke2014nanophotonic} and whispering gallery mode cavities~\cite{Yang2015Advances}. A single-atom cooperativity $C> 500$ for {a} strong single atom-photon coupling can be achieved for microring resonators~\cite{Chang2019Microring}.

{In practice, our protocol is designed for short-distance distributed quantum computation. The length of the fiber channel $L_{\rm fc}$ connecting two neighboring cavities is within the short-fiber limit~\cite{Serafini2006Distributed,cho2008heralded}, ensuring that the interaction time between spatially separated cavities is sufficiently short compared to the cavity mode lifetime~\cite{Chang2020Remote}. The effective interaction between two spatially separated qubits is mediated by the vacuum field, without exciting the atoms or the cavity modes due to the nonresonant couplings in our protocol, except that a single excitation of the normal mode $c_1$ occurs when both qubits decouple from their respective cavity modes. Thus, the presence of fiber attenuation  increases the effective decay rates.}

{Fortunately, the intrinsic loss induced by fiber attenuation can be calculated as $\kappa_{\rm fc}=-c{}{\rm ln}(1-\alpha_l)/(2L_{\rm fc})$~\cite{White2019Cavity}, where $c$ represents the speed of light in the fiber and $\alpha_l$ denotes the single-pass loss of the fiber channel. The impact of the intrinsic loss $\kappa_{\rm fc}$ on the performance of our protocol can be considered to be negligible, given that $\kappa_{\rm fc}$ is approximately $10^{-3}$ of the decay rate of the atomic excited state for a short fiber length of $L_{\rm fc}<1$ m.}

In summary, we   {have} proposed a scheme for implementing a heralded nonlocal CPHASE gate on  spatially separated stationary qubits coupled to different cavities. We can faithfully implement a nonlocal CPHASE gate in a heralded way by dynamically controlling the evolution of a composite system and projecting the auxiliary atom   {onto} a postselected state. We   {have} further showed its application for implementing quantum gates on logical qubits within a DFS. All these distinct  {characteristics} make these quantum gates useful for distributed quantum computation and quantum networks.

\section*{ACKNOWLEDGMENTS}
This work was supported in part by the National Natural Science Foundation of China (Grant No. 11904171).
W.Q. was supported in part by the Incentive Research Project
of RIKEN and acknowledges support of the National Natural  Science  Foundation  of  China  (NSFC)  via  Grant  No.
0401260012. A.M. was supported by the Polish
National Science Centre (NCN) under the Maestro Grant
No. DEC-2019/34/A/ST2/00081.  F.N. is supported in part by
Nippon Telegraph and Telephone Corporation (NTT) Research,
the Japan Science and Technology Agency (JST)
[via the CREST Quantum Frontiers program Grant No. JPMJCR24I2,
the Quantum Leap Flagship Program (Q-LEAP), and the Moonshot R\&D Grant No. JPMJMS2061],
and the Office of Naval Research (ONR) Global (via Grant No. N62909-23-1-2074).

%and the Fundamental Research Funds for the Central Universities (Grant No. 30922010807).

%%%%%%%%%%%%%%%%%%%%%%%%%%%%%%%%%%%%%%%%%%%%%%%%%%%%%%%%%%%%%%%%%%%%%%%%%%%%%%%%%%%%%%%%%%%%%%%%%%%%%%%%%%%%%%%%%%%%%%%%%%%%%%%%%%%%%%%%%%%%%%%%%%%%%%%%%%%%%%%%%%%%%%%%%%%%%%%%%%%
	
%	\nocite{apsrev41Control}  %%%%%  First line
%	\bibliography{ref}

\end{document}